\documentclass[10pt, conference]{IEEEtran}
\IEEEoverridecommandlockouts
\usepackage{url}
\usepackage{cite}
\usepackage{flushend}
\usepackage{makecell}
\usepackage{subfigure}
\usepackage{amsmath,amssymb,amsfonts}
\usepackage{algorithmic}
\usepackage{algorithm}

\usepackage{graphicx}
\usepackage{textcomp}
\usepackage{float}
\usepackage{booktabs}
\usepackage{bm}
\usepackage{float}  
\usepackage{multirow}
\usepackage[normalem]{ulem}
\useunder{\uline}{\ul}{}
\usepackage{xcolor}

\def\BibTeX{{\rm B\kern-.05em{\sc i\kern-.025em b}\kern-.08em
    T\kern-.1667em\lower.7ex\hbox{E}\kern-.125emX}}
\allowdisplaybreaks

\begin{document}

\title{
Communication-Efficient Cooperative Multi-Agent PPO via Regulated Segment Mixture in Internet of Vehicles
\\
}
\author{Xiaoxue Yu, Rongpeng Li, Fei Wang, Chenghui Peng, Chengchao Liang, Zhifeng Zhao, and Honggang Zhang\vspace{-1.5cm}\\
\thanks{This work was supported by the National NSF of China (62071425), the Zhejiang Key R\&D Plan (2022C01093), Huawei Cooperation Project, and the Zhejiang Provincial NSF of China (LR23F010005).

X. Yu, and R. Li are with College of Information Science and Electronic Engineering, Zhejiang University (email: \{sdwhyxx, lirongpeng\}@zju.edu.cn). F. Wang and C. Peng are with Huawei Technologies (email: \{wangfei76, pengchenghui\}@huawei.com). C. Liang is with the School of Communication and Information Engineering, Chongqing University of Posts and Telecommunications (email: chengchaoliang@sce.carleton.ca). Z. Zhao and H. Zhang are with Zhejiang Lab as well as Zhejiang University (email: \{zhaozf, honggangzhang\}@zhejianglab.com).}
}
\maketitle

\begin{abstract}
Multi-Agent Reinforcement Learning (MARL) has become a classic paradigm to solve diverse, intelligent control tasks like autonomous driving in Internet of Vehicles (IoV). However, the widely assumed existence of a central node to implement centralized federated learning-assisted MARL might be impractical in highly dynamic scenarios, and the excessive communication overheads possibly overwhelm the IoV system. Therefore, in this paper, we design a communication
efficient cooperative MARL algorithm, named RSM-MAPPO, to reduce the communication overheads in a fully distributed architecture. In particular, RSM-MAPPO enhances the multi-agent Proximal Policy Optimization (PPO) by incorporating the idea of segment mixture and augmenting multiple model replicas from
received neighboring policy segments. Afterwards, RSM-MAPPO adopts a theory-guided metric to regulate
the selection of contributive replicas to guarantee the policy improvement.
Finally, extensive simulations in a mixed-autonomy traffic control scenario verify the effectiveness of the RSM-MAPPO algorithm.

\end{abstract}
\begin{IEEEkeywords}
Communication-efficient, Multi-agent reinforcement learning, Regulated segment mixture, Internet of vehicles.
\end{IEEEkeywords}

\vspace{-0.8em}
\section{Introduction}\label{sec1}

Internet of Vehicles (IoV) emerges as an effective means to ubiquitously connect vehicles and enhance their self-driving capability (e.g., fleet management and accident avoidance). Typically, in IoV, a Connected Automated Vehicle (CAV) is contingent on Deep Reinforcement Learning (DRL) to solve diverse control tasks \cite{9351818, kreidieh2018dissipating,shi2020efficient}, on top of a formulated Markov Decision Process (MDP). Correspondingly, these CAVs constitute a Multi-Agent Reinforcement Learning (MARL)-empowered system. 
Nevertheless, the direct adoption of Independent Reinforcement Learning (IRL) \cite{tan1993multi} at the CAV, with each one accessible and responsive to a limited partial observation of the global environment, will make MARL suffer from the non-stationarity of the learning environment.
Therefore, communications are generally taken into account as an indispensable ingredient in MARL \cite{xu2021gradient,foerster2016learning,jiang2018learning,kim2021communication}. 
For example, Ref. \cite{xu2021gradient} combines Federate Learning (FL) with IRL, by regarding the aggregation of gradients as the communication, so as to improve the involved homogeneous agents' capability and learning efficiency. Meanwhile, individual observations \cite{foerster2016learning} or intended actions \cite{jiang2018learning,kim2021communication} can also be exchanged on the basis of proper encoding. 
Moreover, Ref. \cite{xu2022trustable} proposes a stigmergy-based trustable policy collaboration scheme by directly mixing the policy parameters.
But the common assumption of an existing central node in these works \cite{foerster2016learning,jiang2018learning,kim2021communication,xu2021gradient,xu2022trustable} might be impractical and underlies potential threat to the stability and timeliness of learning performance in highly dynamic scenarios like IoV.
%
Besides, the frequent information exchange in these works inevitably generates excessive and even exponential communication overheads along with the number of agents, thus possibly overwhelming the IoV system.
In a nutshell, it becomes imperative to design a communication efficient MARL algorithm.

In that regard, there has emerged intense research interest, particularly within the scope of Decentralized Federated Learning (DFL) and supervised learning.
Ref. \cite{barbieri2022communication} puts forward a randomized selection scheme for forwarding subsets of local model parameters to their one-hop neighbors.
Ref. \cite{hu2019decentralized} introduces a segmented gossip approach by synchronizing model segments only, thus significantly splitting the expenditure of communications.
%
However, communication efficient system, which typically adopts a larger communication internal, faces more diverse local model updates, and may get an even worse aggregation model after simple parameter averaging \cite{kairouz2021advances}. Notably, this could be more exacerbated for an on-line IRL framework, since IRL agents need to interact with the environment more frequently than those for supervised learning and the processing of gradually arrived data could amplify the learning discrepancy among multiple agents.
In other words, not all communicated packets will be contributive in MARL and directly adopting the over-simplistic mixture approach as in DFL works \cite{hu2019decentralized,barbieri2022communication} is far from efficiency.
Instead, MARL awaits for a revolutionized mixture method and corresponding metric to regulate the aggregation of exchanged model updates, so as to ensure robust policy improvement.

In this paper, on the basis of Proximal Policy Optimization (PPO) \cite{schulman2017proximal}, one classical policy iteration reinforcement learning algorithm, we tailor a distributed communication-efficient cooperative scheme for IRL-controlled CAVs in IoV, and propose a Regulated Segment Mixture-based Multi-Agent PPO (RSM-MAPPO) algorithm. Compared with existing communication-based MARL works, the key contributions of RSM-MAPPO can be summarized as follows.
\begin{itemize}
    \item RSM-MAPPO implements a communication-efficient MAPPO by incorporating the idea of segment mixture in DFL and augmenting multiple model replicas from received neighboring policy segments.
    \item In order to guarantee the policy improvement during the mixture, a theory-guided metric is developed to regulate the selection of contributive replicas only.
    \item Through extensive simulations in the traffic control scenario, RSM-MAPPO, which operates in a fully distributed manner, could approach the converged performance of centralized FL and IRL \cite{xu2021gradient}, while is significantly superior than direct application of parameters average as in DFL \cite{hu2019decentralized,barbieri2022communication}, thus verifying its effectiveness. 
\end{itemize}

The remainder of this paper is organized as follows.
We introduce the system model and formulate the problem in Section \ref{sec2}.
Afterwards, we elaborate on the details of the proposed RSM-MAPPO algorithm in Section \ref{sec3}.
In Section \ref{sec4}, we present the simulation settings and discuss the experimental results.
Finally, Section \ref{sec5} concludes this paper.

\section{System Model and Problem Formulation}\label{sec2}
\begin{figure}[]
\centering 
\includegraphics[scale = 1.7]{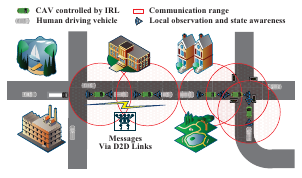}
\vspace{-1.5em}
\caption{Illustration of MARL in autonomous driving. }
\label{autodriv}
\vspace{-1.8em}
\end{figure} 
Beforehand, we summarize the main notations in Table \ref{notions}.
\vspace{-0.5em}
\subsection{System Model}
As illustrated in Fig. \ref{autodriv}, we primarily consider an IoV scenario consisting of $N$ CAVs (i.e., PPO-empowered agents)\footnote{In this paper, we assume the terminologies ``CAV'' and ``agent'' are inter-changeable.} alongside some human-driving vehicles.
Specifically, at each time-step $t$, agent $i$ senses partial status $s_t^{(i)}$ (e.g., the speed and positions of neighboring vehicles) of the IoV environment, and then selects an action $a_t^{(i)}\!\in\!\mathcal{A}$ according to its local policy $\pi^{(i)}$ parameterized by $\theta^{(i)}$. Afterwards, an individual reward $r_t^{(i)}\in \mathcal{R}$ will be obtained, with the state transferring into $s_{t+1}^{(i)}$. Correspondingly, a sequential Markov state transition $\phi_t^{(i)} = \langle s_t^{(i)},a_t^{(i)},r_t^{(i)},s_{t+1}^{(i)} \rangle$ can be stored. 
In this paper, the MAPPO learning encompasses an independent local learning phase and a communication-assisted mixing phase. Generally, in the first phase, after running a policy for $T$ time-steps (far less than the length of an epoch, which equals multiple $T$), we can obtain a mini-batch $\Phi$ of collected samples for iterations of local model updates. 
Subsequently, in the second phase, each agent $i$ interacts with its one-hop neighbors $\Omega_i$ within its communications range directly (e.g., via Device-to-Device (D2D) channels), so as to reduce the behavioral localities of IRL and improve their cooperation efficiency.
\begin{table}[tbp]
    \renewcommand\arraystretch{1}
    \normalsize
    \centering
    \caption{Major notations used in the paper.}
    \vspace{-0.6em}
    \label{n_table}
    \scalebox{0.7}{
    \begin{tabular}{ll}
    \hline
    Notation & Definition \\ \hline
    $s_t^{(i)},a_t^{(i)},r^{(i)}_t$ & Local state, individual action and reward of agent $i$ at time step $t$\\
    $\pi$, $\theta$ & Current target policy and its parameters\\
    $\tilde{\pi}$, $\tilde{\theta}$ & The referential target policy and it parameters \\
    $\Omega_i$ & Set of one-hop neighbors within the communication range of agent $i$\\
    $\alpha$ & Mixture metric of current parameters and referential parameters \\
    $\theta_{\text{mix}}$ & Mixed policy parameters \\
    $p,P$ & Index of segments, $p=1,2,\cdots,P$\\
    $\kappa$ & Number of model replicas \\
    $\tau U$ & Communication interval given $U$ local iterations\\
    $\upsilon$ & Size of the policy parameters \\
    $\psi$ & Communication consumption until convergence of the IRL model \\\hline
    \end{tabular}
}
    \label{notions}
    \vspace{-0.6cm}
\end{table}

Algorithmically, we adopt a sample-efficient standard PPO setting in the local learning phase, which leverages two different policies (i.e., behavior policy $\pi_{\theta_{\text{old}}}$
\footnote{Hereafter, for simplicity of representation, we omit the superscript $(i)$ under cases where the mentioned procedure applies for any agent.} 
for sample collection and target policy $\pi_\theta$ for online optimization) instead of the same policy in classical REINFORCE. Every $U$ local iterations, the parameters of the target policy will be copied to those of the behavior policy. In addition, PPO implements importance sampling-based optimization using all past experiences via an adjustable ratio $\lambda_t= \frac{\pi_{\theta}(a_t|s_t)}{\pi_{\theta_\text{old}}(a_t|s_t)}$ without leading to destructively large policy updates. 
Thus, the actor network's loss function 
is expressed as
\begin{align}
\vspace{-0.5em}
    L(\theta)\! &=\! -\!\mathbb{E}_{t}\!\Big[\!\min\![\lambda_tA^{\pi_{\!\theta_{\!\text{old}}}}_{t}\!,\!\text{clip}(\!\lambda_t,\!1-\epsilon,\!1+\epsilon)A^{\pi_{\!\theta_{\!\text{old}}}}_{t}\!]\!+\!\beta\!H\!\left[\pi_\theta(s_t)\right]\!\Big]\nonumber
\vspace{-0.6em}
\end{align}
where the operator $\mathbb{E}_t(\cdot)$ indicates a $T$-length empirical average over a batch of samples with $t \in [0,T-1]$, and the entropy function $H(\cdot)$ ensures sufficient exploration, while $\beta$ is a hyperparameter to reflect the relative importance of entropy. Besides, the function $\text{clip}(\cdot, 1-\epsilon, 1+ \epsilon)$ aims to penalize over-large policy changes and clips the ratio into $[1\!-\!\epsilon,1\!+\!\epsilon]$, where $\epsilon$ is a hyperparameter. 
Furthermore, $A^{\pi_{\theta_{\text{old}}}}_{t} = \delta_t+\gamma\delta_{t+1}+\cdots+\gamma^{T-t-1}\delta_{T-1}$ is an estimator of the advantage function at timestep $t$, where $\delta_t=r_t+\gamma V_\omega(s_{t+1})-V_\omega(s_t)$, and $\gamma$ denotes a discount factor. 
%
Along with the update of policy $\pi_{\theta}$, $V_\omega(s_t)$ parameterized by $\omega$ is estimated by another critic network in terms of Mean Squared Error (MSE) loss
\vspace{-0.2em}
\begin{equation}
\vspace{-0.6em}
    L(\omega)=\mathbb{E}_t\left[(V_\omega(s_t)-V^{\text{targ}}_t)^2\right] \label{eq:v_loss}\nonumber
\end{equation}
where $V^{\text{targ}}_t$ is the target value equals $\sum_{i=0}^{T-t-1}\!\!\gamma^{i}r_{t+i}\!+\!\gamma^{T-t}V_\omega(s_{T})$.
\noindent 
\begin{figure*}[tbp]
\subfigcapskip = -10pt
\begin{center}
    \subfigure[Independent local learning phase]
    {\includegraphics[scale =0.31]{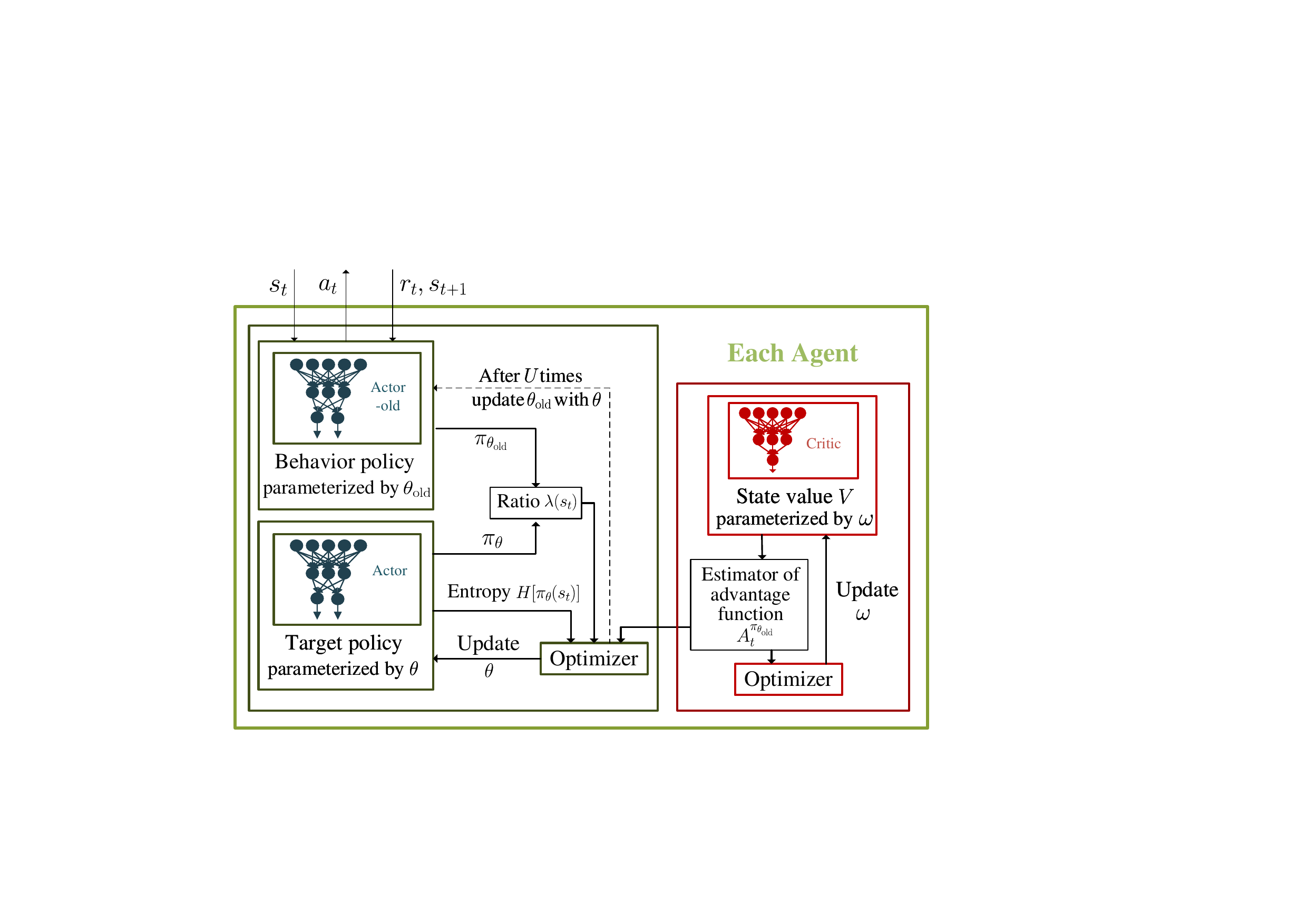}}
    \subfigure[Communication-assisted mixing phase]
    {\includegraphics[scale =0.4]{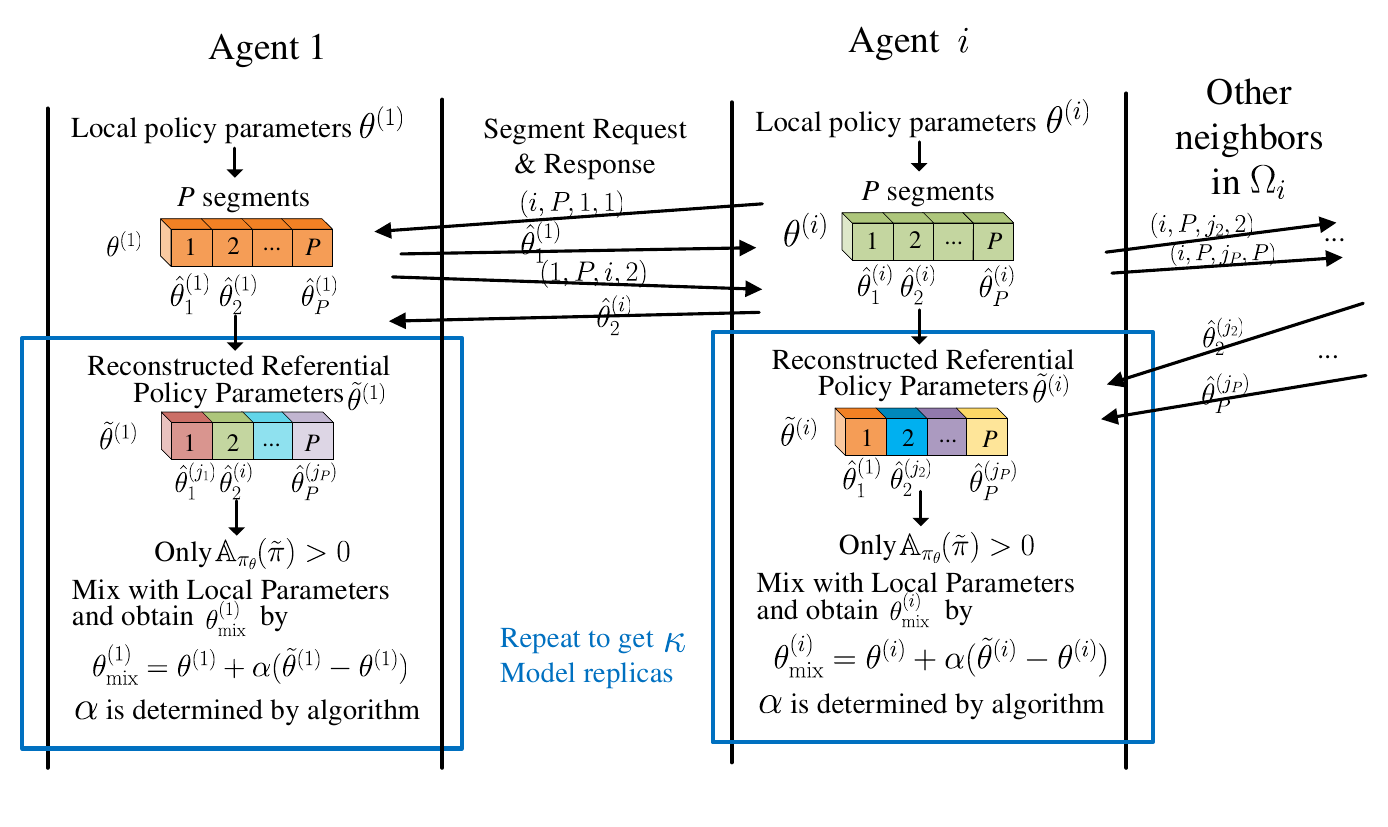}}
\vspace{-0.3em}
\caption{The illustration of RSM-MAPPO implementation.}  
\vspace{-0.6cm}
\label{fig:framework}
\end{center}
\end{figure*} 
\noindent At local iteration $k$, the parameter update follows a standard Stochastic Gradient Descent (SGD) as
\begin{equation}
\vspace{-0.5em}
    \theta_{k+1}^{(i)} = \theta_k^{(i)}-\eta_a \nabla L(\theta_k^{(i)})\label{eq:actor_iteration}
\end{equation}
\begin{equation}
\vspace{-0.2em}
    \omega_{k+1}^{(i)} = \omega_k^{(i)}-\eta_c \nabla L(\omega_k^{(i)})\label{eq:critic_iteration}
\vspace{-0.2em}
\end{equation}
where $\eta_a$ and $\eta_c$ are the learning rate of the actor network and the critic network respectively. 

Upon every $\tau U$ local iterations (i.e., $\tau$ times of copying parameters from $\pi_{\theta_{\text{old}}}$ to $\pi_{\theta}$), the communications among neighboring agents starts. 
Considering the possible communication bandwidth or delay restriction between agents in real-world facilities, we assume messages transmitted by agents are limited to policy parameters $\theta$, and agents could develop different means to derive a referential policy parameterized by $\tilde{\theta}\!=\!f({\theta}^{(1)},\cdots,{\theta}^{(j)},\cdots)$ by exploiting the parameters from neighboring agents $\forall j \in \Omega_i$. 
For example, agent $i$ could compute a referential policy $\tilde{\pi}^{(i)}$ parameterized by $\tilde{\theta}^{(i)}$ based on received parameters $\theta^{(j)}$ from $j \in \Omega_i$,
and directly mix neural network parameters distributedly as
\vspace{-0.2em}
\begin{equation}
\theta_\text{mix}^{(i)}=\theta^{(i)} +\alpha(\tilde{\theta}^{(i)}-\theta^{(i)})
    \label{eq:mix}
\vspace{-0.2em}
\end{equation}
where $\alpha \in [0,1]$ is a mixture metric. Taking general parameter average mixture method in DRL-Ave \cite{xu2022trustable} as an example, $\tilde{\theta}^{(i)}$ is computed as $\tilde{\theta}^{(i)}\!=\!\sum\nolimits_{j\in\Omega_i}\!{\theta}^{(j)}$ and $\alpha\!=\!1\!-\!1/|\Omega_i|$, which is influenced by the number of neighbors involved. Subsequently, for each agent $i$, $\theta^{(i)}$ and $\theta_{\text{old}}^{(i)}$ get aligned with $\theta_{\text{mix}}^{(i)}$.

\subsection{Problem Formulation}

This paper primarily targets the communication assisted mixing phase. 
Intuitively, an effective mixture means could better leverage the exchanged parameters to yield a superior target policy and thus benefit the learning in terms of the rewards along with the learning trajectory. In other words, the reward could be a function of the mixed policy parameters $\Theta = \{\theta_\text{mix}^{(1)},\cdots, \theta_\text{mix}^{(N)} \}$ and $\alpha$. However, it remains little investigated on the feasible means to mix the exchanged parameters (or their partial segments), though it vitally affects both the communication overheads and learning performance. Therefore, by optimizing both $f$ and $\alpha$, we mainly focus on reducing the communication expenditure while maintain an acceptable cumulative rewards, that is,
\begin{align}
\vspace{-0.5em}
& \min_{f,\alpha} c(\upsilon,f)\nonumber \\
s.t.\quad 
& \sum\nolimits_t r_t(\Theta, \alpha) \geq r^{\text{thre}} \nonumber \\
& \Theta \leftarrow \{\theta_{\text{mix},k}^{(1)},\cdots, \theta_{\text{mix},k}^{(N)} \}\\
& \theta_{\text{mix},k}^{(i)}=\theta_k^{(i)} +\alpha(\tilde{\theta}_k^{(i)}-\theta_k^{(i)}), &  \forall i \in \{1,\cdots, N\} \nonumber \\
& \tilde{\theta}_k^{(i)} = f({\theta}_k^{(1)},\cdots,{\theta}_k^{(j)},\cdots), & \forall k\text{ mod } \tau U =0, j \in \Omega_i \nonumber 
\end{align}
where $r^{\text{thre}}$ denotes the required minimum cumulative rewards, and $v$ indicates the size of policy parameters. Furthermore, $c(v,f)$ denotes the communication expenditure, which is governed by the mixture function $f$. For example, for the whole policy parameters transmission among all agents \cite{xu2022trustable}, the total communication cost per round is $c(\upsilon,f)\!=\!N\!\times\!(N-1)\!\times\!\upsilon$. Apparently, the communication cost $c(\upsilon,f)$ can be significantly reduced, if $f$ could rely on fewer agents with reduced communication frequency. However, such a naive design possibly mitigates the positive effect of collaboration as well. Therefore, it is worthwhile to resort to a more comprehensive design of $f$ and $\alpha$ to calibrate the communicating agents and content as well as regulate the mixture means, so as to provide a guarantee of performance improvement.

\section{MAPPO with Regulated Segment Mixture}\label{sec3} 
In this section, as shown in Fig. \ref{fig:framework}, we present the design of RSM-MAPPO, which reduces the communication overheads on the basis of not much learning performance sacrifice.
\subsection{Algorithm Design}\label{4B1}
Consistent with the standard PPO as in Section \ref{sec2}, agents in RSM-MAPPO undergo the same local iteration process. Meanwhile, for the communication-assisted mixing phase, RSM-MAPPO typically entails segment request \& response, model replica building, and parameter mixture with theory-established performance improvement.


\subsubsection{Segment Request \& Response}

Inspired by segmented pulling synchronization in DFL \cite{hegedHus2019gossip}, we develop and perform a segment request \& response procedure, which allows the agent to request different parts of its policy parameters from different neighbors and rebuild a mixed referential policy for aggregation.
Specifically, for every communication round, each agent $i$ breaks its policy parameters $\theta^{(i)}$ into $P$ ($P \leq \vert \Omega_i\vert$) non-overlapping segments $\hat{\theta}^{(i)}_1,\hat{\theta}^{(i)}_2,\cdots,\hat{\theta}^{(i)}_P$ as
\vspace{-0.5em}
\begin{equation}
\vspace{-0.2cm}
    \theta^{(i)} =(\hat{\theta}^{(i)}_1,\hat{\theta}^{(i)}_2,\cdots,\hat{\theta}^{(i)}_P)
\end{equation}
Notably, available segmentation strategies include, but not limited to, dividing the policy parameters according to the neural network layers, the amount of samples each agent collected, the size of total parameters, etc.
Here, we consider the most intuitive parameters uniform partition to clarify this process.
And for each segment $p=1,\cdots, P$, agent $i$ randomly selects a target agent (without replacement) from its neighbors (i.e., $j_p \in \Omega_i$) to send segment request $(i,P,j_p,p)$, which indicates the agent $i$ who initiates the request and its total segment number $P$, the target agent $j_p$ that will receive the request and break its own policy parameters $\theta^{(j_p)}$ into also $P$ segments, return the corresponding requested segment $\hat\theta^{(j_p)}_p$ in response according to the identifier $p$. 
It should be stressed that in order to reduce the complexity and  facilitate the implementation, we only discuss the case as in Fig. \ref{fig:framework} that $P$ is the same constant for all agents and is not greater than $\max_i\vert \Omega_i\vert,\forall i$. Then, agent $i$ could reconstruct a referential policy based on all of the fetched segments, that is, 
\begin{equation}\label{reconstruct}
    \tilde{\theta}^{(i)} = (\hat{\theta}^{(j_1)}_1,\hat{\theta}_2^{(j_2)},\cdots,\hat{\theta}_P^{(j_P)})
\vspace{-0.1cm}
\end{equation}
This step, which can be conveniently performed in parallel to make full use of the bandwidth, contributes to avoiding the model staleness, since one reconstructed model consists of different agents' latest update policy segments, thus propagating more agents' local updates through the whole system. 

\subsubsection{Model Replica Building}\
\addtolength{\topmargin}{0.05in}
\begin{algorithm}[tbp]\small
  \caption{Communication-assisted mixing phase of RSM-MAPPO Alogrithm}
  \label{my alg}
  \begin{algorithmic}[1]
    \REQUIRE {the target policy's parameters ${\theta}^{(i)}$ for $i=1,2,\cdots,N$; number of samples to estimate policy advantage $M$; number of samples to evaluate FIM $K$; number of replica $\kappa$; number of segment $P$.}
    \ENSURE {${\theta_\text{mix}}^{(i)}$ for $i=1,2,\cdots,N$;}
        \STATE {\textbf{Each agent $i$ executes:}}
            \FOR {each replica $u=1,2,\cdots,\kappa$}
                \STATE Send $P$ pulling request $(i,P,j_p,p)$ to nearby collaborators in $\Omega_i $, and receive $\hat{\theta}^{(j_p)}_p$ to reconstruct $\tilde{\theta}$ as \eqref{reconstruct}.
                \STATE Randomly select $M$ samples from the replay buffer of agent $i$ under the behavior policy $\pi_{\theta_\text{old}}$ to estimate $\mathbb{A}_{\pi_\theta}(\tilde{\pi})$ according to \eqref{advance};
                \IF {$\mathbb{A}_{\pi_\theta}(\tilde{\pi})>0$}
                    \STATE Randomly select $K$ samples from the replay buffer of agent $i$ to evaluate $G(\theta^{(i)})$ according to \eqref{FIM}. 
                    \STATE Get the upper bound of $\alpha$ according to Theorem \ref{theorem}.
                    \STATE Make the mixture metric $\alpha$ less than the calculated upper bound, and update $\theta^{(i)}$ by \eqref{eq:mix}.
                \ENDIF
                \ENDFOR
    \RETURN{the referential policy's parameters ${\theta_\text{mix}}^{(i)}$ for $i=1,2,\cdots,N$.}
  \end{algorithmic}
\end{algorithm}
As it is difficult to bound the staleness of model updates, we adopt the concept of model replica into RSM-MAPPO, so as to further accelerate the propagation and ensure the model quality. Specifically, each agent $i$ repeats the process of segment request and response for $\kappa$ times, thus reconstructing $\kappa$ distinctive model replicas. 

\subsubsection{Parameter Mixture with Theory-Established Performance Improvement}\ 
As the policy performance may vary significantly due to the differences in training samples of multiple agents, there might emerge some reconstructed model replicas degrading the learning performance, and a direct application of averaging mixture method in Section \ref{sec2} possibly makes the aforementioned procedures in vain. Instead, based on our previous works \cite{xu2022trustable}, we derive the following mixture metric to justify the effectiveness of a model replica and only select the contributive ones. Beforehand, we give the following useful theorem \cite{xu2022trustable}.

\newtheorem{theorem}{Theorem}
\begin{theorem}\label{theorem}
For a PPO agent with a current target policy $\pi_\theta$ and a referential policy $\tilde{\pi}$ parameterized by $\theta$ and $\tilde{\theta}$ respectively, if
\begin{enumerate}
    \item $\mathbb{A}_{\pi_\theta}(\tilde{\pi})> 0 $
    \item $
   0<\alpha<\left[ 2\left( \frac{\mathbb{A}_{\pi_\theta}(\tilde{\pi})}{C}\right)^{\frac{1}{2}} / \left[ (\tilde{\theta}-\theta)^T G(\theta) (\tilde{\theta}-\theta) \right]\right]^{\frac{1}{2}}$
\end{enumerate}
the cumulative rewards are guaranteed to be improved through updating $\theta$ to $\tilde{\theta}$ according to \eqref{eq:mix}.
Notably, $C = \frac{2\varepsilon\gamma}{(1-\gamma)^2}$, $\varepsilon = \max_{s_t}\max_{a_t}|\delta_t|$. 
$\mathbb{A}_{\pi_{\theta}}(\tilde{\pi})$ is defined as the expectation of the advantage function along the $\tilde{\pi}$-yielded learning trajectory, and can be approximated as the expectation of the multiplication of policy gain $\tilde{\pi}-\pi_\theta$ and 1-step advantage function $\delta_t$ along with the $\pi_{\theta_{\text{old}}}$-yielded learning trajectory. Meanwhile, $G(\theta)$ is the Fisher Information Matrix (FIM) of policy parameters $\theta$.
\end{theorem}

Based on Theorem \ref{theorem}, we can verify the contribution of a model replica by computing $\mathbb{A}_{\pi_\theta}(\tilde{\pi})$, and get the upper bound of $\alpha$ by further computing $G(\theta)$ from Monte-Carlo simulations, that is,
\begin{align}
   &\mathbb{A}_{\pi_\theta}(\tilde{\pi}) \approx \mathbb{E}_{t}\!\!\left[\frac{\tilde{\pi}(a_t|s_t)\!-\!\pi_\theta(a_t|s_t)}{\pi_{\theta_\text{old}}(a_t|s_t)}\right]\delta_t
    \label{advance}\\
    &G(\theta) \approx \mathbb{E}_{t}\!\left[ \left(\frac{\partial \log \pi_\theta(a_t|s_t)}{\partial \theta} \right)\!\!\left(\frac{\partial \log \pi_\theta(a_t|s_t)}{\partial \theta}\right)^{\top}\right]\label{FIM}
\end{align}
%

\noindent Afterwards, we can select and mix those model replicas with positive $\mathbb{A}_{\pi_\theta}(\tilde{\pi})$, which means the agent can benefit from mixing its policy parameters $\pi_\theta$  with the reconstructed referential policy $\tilde{\pi}$. More aggressively, it is also feasible to merge the model replica with the largest $\mathbb{A}_{\pi_\theta}(\tilde{\pi})$ only. Besides, since $G(\theta)$ is a positive definite matrix, the mixture metric $\alpha$ will enlarge with the increase of $\mathbb{A}_{\pi_\theta}(\tilde{\pi})$. Thus, a better referential policy contributes to faster learning as well.

Finally, we summarize the details of RSM-MAPPO in Algorithm \ref{my alg}.

\subsection{Discussions of Communication overheads}\label{4B2}

In regards to the communication overheads per segment request, RSM-MAPPO costs $\upsilon/P$ amount of data transmission via D2D communications. Therefore, the total amount of communications overheads per round equals $N\times \upsilon$, which is $N-1$ times less than that in \cite{xu2022trustable}. Meanwhile, by simultaneously requesting $P$ agents, it benefits the sufficient use of the bandwidth and enhances the capability to overcome possible channel degradation. On the other hand, for cases with $\kappa$ model replicas, the communication overheads per round turns to $N\times\kappa\times\upsilon$, which is $\frac{N-1}{\kappa}$ times less than that in \cite{xu2022trustable}, but improves the learning performance.

\section{Experimental Results and Discussions}\label{sec4}

\subsection{Experimental Settings}\label{4B}
\begin{table}[tbp]
    \centering
    \caption{System Parameters.}
    \label{tab:parameters}
    \begin{tabular}{l|c|c}
    \toprule
    Parameters  & Symbol & Value\\ \midrule
    Total time-steps of an epoch & $E$ & $1500$ \\
    Number of timesteps for a mini-batch & $T$ & $250$ \\
    Number of PPO iterations in a mini-batch & $U$ & $3$ \\
    Number of samples to evaluate
FIM &$K$ & $50 $ \\
Number of samples
to estimate $\mathbb{A}_{\pi_\theta}(\tilde{\pi})$ &$M$ & $200$ \\
    Learning rate of actor network &$\eta_a$ & $2.5 \cdot 10^{-5}$ \\
    Learning rate of critic network&$\eta_c$ & $5 \cdot 10^{-5}$ \\
    Discount factor &$\gamma$ & $0.9$ \\ 
    Entropy coefficient &$\beta$ & $0.01$ \\ 
    Coefficient of communication internals  & $\tau$ & $1$\\
    Number of segments & $P$ & 4\\
    Number of model replicas & $\kappa$ & 2\\
    \bottomrule
    \end{tabular}
    \vspace{-0.5cm}
\end{table}

In this part, we consider the simulation scenario ``Figure Eight'', a widely-used mixed-autonomy traffic control scenario, to testify the performance (i.e., maximizing the cumulative rewards) of DRL \cite{vinitsky2018benchmarks}. 
There are totally $14$ vehicles running circularly along a one-way lane that resembles the shape of figure ``8''. These include $5$ emulated human-driving vehicles, controlled by Simulation of Urban MObility (SUMO) with a microscopic car-following model named Intelligent Driver Model (IDM)\cite{treiber2000congested}, and $9$ IRL-controlled CAVs, which simultaneously maintain dedicated links to update their parameters through the D2D channel. 
Besides, the scenario is modified to assign the limited partial-observation of global environment as the state of each vehicle, including the position and speed of its own, the vehicle ahead and behind. 
Meanwhile, each CAV's action is a continuous variable representing the speed acceleration or deceleration normalized between $[-1,1]$. In order to reduce the occurrence of collisions and promote the traffic flow to the maximum desired speed, the reward function is $\mathcal{R}=\frac{\max\{\Vert \bm{v}_\text{de} \Vert-\Vert \bm{v}_\text{ac}-\bm{v}_\text{de}\Vert,0\}}{\Vert \bm{v}_\text{de} \Vert}$ 
\footnote{Notably, we assume complete knowledge of individual vehicle speeds at each vehicle here. Beyond the scope of this paper, some value-decomposition method like \cite{xiao_stochastic_2023} can be further leveraged to derive a decomposed reward, so as to loosen such a strict requirement.}, 
where $\bm{v}_\text{de}\in\mathbb{R}^{14}$ and $\bm{v}_\text{ac}\in\mathbb{R}^{14}$ represent the desired velocity and actual velocity of all vehicles in the system respectively. 
In addition, the current epoch will be terminated once a collision occurs. We perform tests every $10$ epochs and take the average of accumulated rewards in a testing epoch as average reward. Besides, all results are produced using the average of $5$ repetitions.
The main parameters used in simulations are listed in Table \ref{tab:parameters}. 
\subsection{Evaluation Metrics}\label{5Eval}
Besides average reward, we also adopt other metrics to extensively evaluate communication efficiency of RSM-MAPPO. 
\begin{itemize}
    \item We use $\rho_\text{total}$ to represent the total number of reconstructed referential policy $\tilde{\pi}$ (i.e., all model replicas) until convergence, that is, the inflexion point of average reward curve. 
    Moreover, we use $\rho_\text{ef}$ to indicate the number of effectively reconstructed referential policy (i.e., contributive model replicas selected to mix). Correspondingly, we further define the ratio $\rho_r = \rho_\text{ef}/\rho_\text{total}$ to reflect the utilization rate of reconstructed policies.
    \item We use $\psi$ to indicate the communication overheads (in terms of $\upsilon$) until convergence. Mathematically, as the number of communication rounds until convergence can be computed as $C_0 = \rho_\text{total} / (N\times\kappa)$, the communication overheads equal $\psi = C_0\times (N\times\kappa\times\upsilon) = \rho_\text{total} \times \upsilon $. 
\end{itemize}
Intuitively, the average reward and communication efficiency will be determined by the function design of $f$, as well as the number of model segments and the model replica (i.e., $P$ and $\kappa$). Besides, the communication overheads are also affected by the coefficient of communication intervals $\tau $. 
\subsection{Simulation Results}\label{5C}

\begin{table*}[tbp]
    \centering
    \caption{Average reward \& communication efficiency of RSM-MAPPO with respect to the metrics in Section \ref{5Eval}.}
    \label{tab:my-table_1}
    
    \begin{tabular}{c|c|c|c|c|c|c|c|c}
\toprule
Method                            & $\tau$ & $P$           &     $\kappa$       & Average Reward    & $\rho_\text{total}$ & $\psi$ & $\rho_\text{ef}$ &  $\rho_r$\\ \midrule
\multirow{2}{*}{Average   mixture} & \multirow{7}{*}{1} & 3                  & \multirow{2}{*}{2} & 0.1987  & $2.4948\cdot10^4$                                                                                                        & $2.4948\cdot10^4\times\upsilon$                      & $2.4948\cdot10^4$                                                                                                                  & 100\%                                                                                                       \\ \cline{3-3} \cline{5-9} 
                                  &                    & 4                  &                    & 0.1934 & $2.7108\cdot10^4$                                                                                                        & $2.7108\cdot10^4\times\upsilon$                      & $2.7108\cdot10^4$                                                                                                                  & 100\%                                                                                                       \\ \cline{1-1} \cline{3-9} 
\multirow{8}{*}{RSM-MAPPO}     &                    & \multirow{4}{*}{4} & 1                  & 0.2121 & $1.3010\cdot10^4$                                                                                                        & $1.3010\cdot10^4\times\upsilon$                      & $1.3010\cdot10^3$                                                                                                                   & 40.402\%                                                                                                    \\ \cline{4-9} 
                                  &                    &                    & 2                  & 0.2116 & $2.2788\cdot10^4$                                                                                                        & $2.2788\cdot10^4\times\upsilon$                      & $9.110\cdot10^3$                                                                                                                   & 39.977\%                                                                                                    \\ \cline{4-9} 
                                  &                    &                    & 4                  & 0.2137 & $5.4216\cdot10^4$                                                                                                        & $5.4216\cdot10^4\times\upsilon$                      & $2.0127\cdot10^4$                                                                                                                  & 37.124\%                                                                                                    \\ \cline{4-9} 
                                  &                    &                    & 8                  & 0.2128 & $8.2512\cdot10^4$                                                                                                        & $8.2512\cdot10^4\times\upsilon$                      & $3.0798\cdot10^4$                                                                                                                  & 37.325\%                                                                                                    \\ \cline{2-9} 
                                  & 5                  & \multirow{3}{*}{3} & \multirow{3}{*}{2} & 0.2110 & $6.498\cdot10^3$                                                                                                         & $1.300\cdot10^4\times\upsilon$                       & $2.711\cdot10^4$                                                                                                                   & 41.721\%                                                                                                    \\ \cline{2-2} \cline{5-9} 
                                  & 10                 &                    &                    & 0.2080 & $3.240\cdot10^3$                                                                                                         & $6.480\cdot10^3\times\upsilon$                       & $1.333\cdot10^3$                                                                                                                   & 41.141\%                                                                                                    \\ \cline{2-2} \cline{5-9} 
                                  & 15                 &                    &                    & 0.2100 & $2.376\cdot10^3$                                                                                                         & $4.752\cdot10^3\times\upsilon$                       & $1.016\cdot10^3$                                                                                                                   & 42.761\%                                                                                                    \\ \bottomrule
\end{tabular}
\vspace{-0.4cm}
    \end{table*}

\begin{figure}[tbp]
\vspace{0.1in}
    \centering
    \includegraphics[scale=0.48]{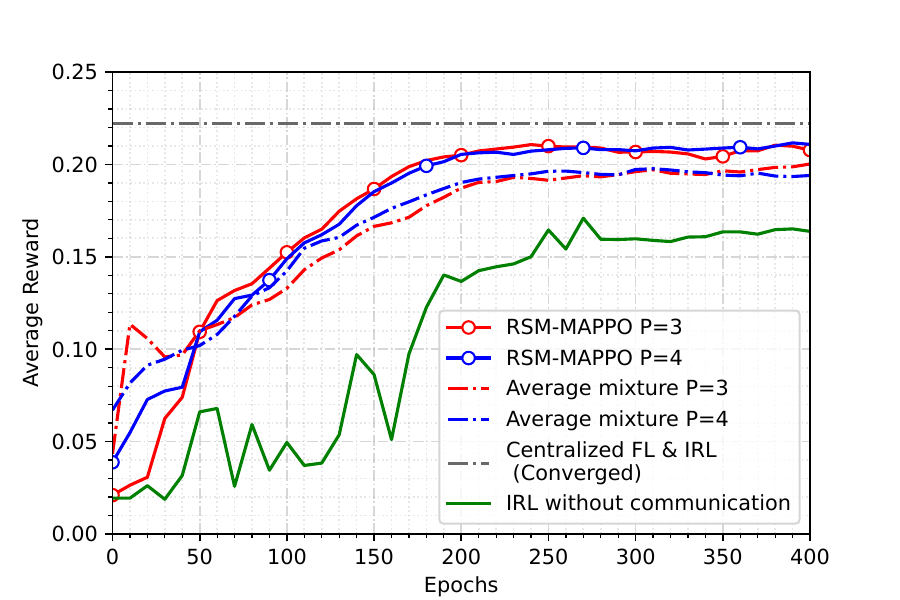} 
    \vspace{-0.3cm}
\caption{Average reward under different methods.}
\vspace{-0.2cm}
\label{fig:S}
\end{figure}

\begin{figure}[tbp]
\vspace{-0.2em}
    \centering
    \includegraphics[scale=0.47]{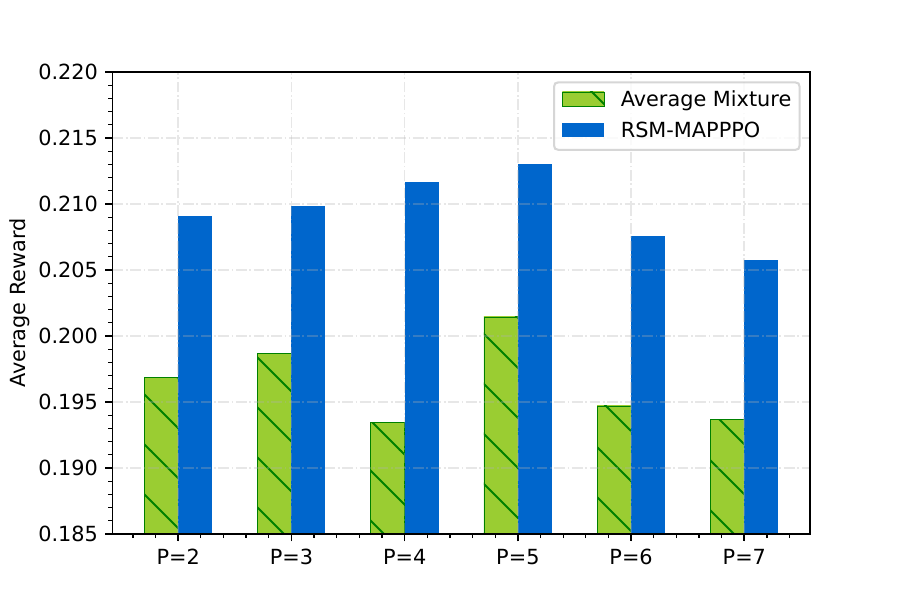} 
    \vspace{-0.3cm}
\caption{Average reward of average mixture\&RSM-MAPPO with different $P$.}
\vspace{-0.3cm}
\label{fig:different P}
\end{figure}

\begin{figure}[tp]
    \centering
    \includegraphics[scale=0.48]{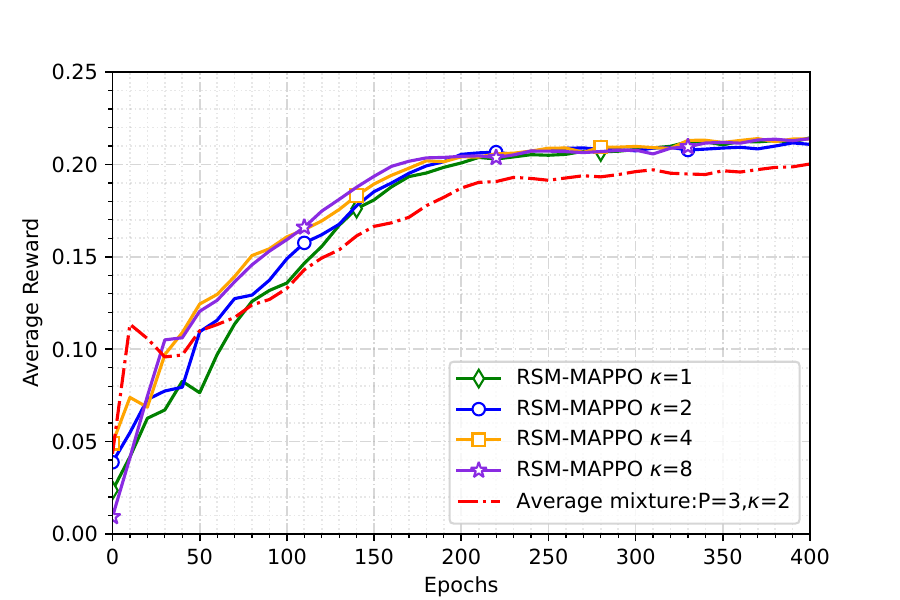}
    \vspace{-0.3cm}
\caption{Average reward of RSM-MAPPO with different $\kappa$.} 
\label{fig:R} 
\vspace{-1.4em}
\end{figure} 

Fig. \ref{fig:S} first examines the average reward with RSM-MAPPO $\kappa=2$. 
It can be observed from Fig. \ref{fig:S}, the curve of IRL without communications implies that simply extending IRL to multi-agent scenarios without any cooperation cannot solve complex tasks.
As a comparison, other methods adding communications among agents can clearly boost the learning performance in terms of training efficiency and stability.
Besides, we use the well trained model under combination of centralized FL and IRL \cite{xu2021gradient} as the optimal baseline. It can be seen that our RSM-MAPPO which is performed in a fully distributed training process, approaches the converged performance of the centralized method. Meanwhile, compared with simply average mixture, which directly takes the average of all replicas, 
RSM-MAPPO also yields superior converged average reward, which can also be further validated in Fig. \ref{fig:different P}. On the other hand, Fig. \ref{fig:S} shows that partitioning the model into different segments (i.e., different $P$) leads to similar convergence trend. However, the converged average reward is relevant to the exact value of $P$, as demonstrated in Fig. \ref{fig:different P}, 
which investigates this influence. Specifically, the final converged average reward becomes higher at first with the increase of $P$ from $2$ to $5$, but then decreases when $P=6$ and $P=7$. 
The performance degradation in the latter cases is because that the aggregation target of reconstructed policy parameters for an over-large $P$ is mottled and loses integrality. 

In addition, Fig. \ref{fig:R} studies the impact of the number of model replicas $\kappa$ on average reward. As shown in Fig. \ref{fig:R}, an increase of $\kappa$ could accelerate the convergence of training process, without apparent influences on the converged average reward. 
On the other hand, the improvement in the convergence rate comes at the cost of increased communication overheads $\psi$, which is listed in Table \ref{tab:my-table_1}. With the increase of $\kappa$, both $\psi$ and $\rho_{\text{ef}}$ increase, but the ratio $\rho_r$ does not increase scalely. Therefore, the trade-off between convergence rate improvement and communication overheads need to be further considered. 
Furthermore, we testify the performance under different values of communication interval $\tau$. Due to the space limitation, the results along with the detailed comparison of the communication efficiency is summarized in Table \ref{tab:my-table_1}. The communication overheads $\psi$ are reduced by $\tau$ times compared with $\rho_\text{total}\times\upsilon$, resulting into higher $\rho_r$ for a larger $\tau$.

\section{Conclusions}\label{sec5}

In this paper, we have proposed a communication-efficient algorithm RSM-MAPPO to deal with the excessive communication overheads among distributed MARL.
%
By delving into the policy parameter mixture function, RSM-MAPPO has provided a novel means to leverage and boost the effectiveness of distributed multi-agent collaboration. In particular, RSM-MAPPO has successfully transformed the classical means of complete parameter exchange into segment-based request and response, which significantly facilitates the construction of multiple model replicas and simultaneously captures enhanced learning diversity. Moreover, in order to avoid performance-harmful parameter mixture, RSM-MAPPO has leveraged a theory-established regulated mixture metric to select the contributive replicas with positive relative policy advantage only. Finally, extensive simulations have demonstrated the effectiveness of this design. 
In the future, we will extend this regulated segment mixture paradigm to more RL algorithms to verify its generalization. 

\bibliographystyle{IEEEtran}
\bibliography{reference}

\begin{thebibliography}{10}
\providecommand{\url}[1]{#1}
\csname url@samestyle\endcsname
\providecommand{\newblock}{\relax}
\providecommand{\bibinfo}[2]{#2}
\providecommand{\BIBentrySTDinterwordspacing}{\spaceskip=0pt\relax}
\providecommand{\BIBentryALTinterwordstretchfactor}{4}
\providecommand{\BIBentryALTinterwordspacing}{\spaceskip=\fontdimen2\font plus
\BIBentryALTinterwordstretchfactor\fontdimen3\font minus
  \fontdimen4\font\relax}
\providecommand{\BIBforeignlanguage}[2]{{%
\expandafter\ifx\csname l@#1\endcsname\relax
\typeout{** WARNING: IEEEtran.bst: No hyphenation pattern has been}%
\typeout{** loaded for the language `#1'. Using the pattern for}%
\typeout{** the default language instead.}%
\else
\language=\csname l@#1\endcsname
\fi
#2}}
\providecommand{\BIBdecl}{\relax}
\BIBdecl

\bibitem{9351818}
B.~R. Kiran\emph{,~et~al.}, ``Deep {R}einforcement {L}earning for {A}utonomous
  {D}riving: {A} {S}urvey,'' \emph{IEEE Trans. Intell. Transp. Syst.}, vol.~23,
  no.~6, pp. 4909--4926, Jun. 2022.

\bibitem{kreidieh2018dissipating}
A.~R. Kreidieh\emph{,~et~al.}, ``Dissipating stop-and-go waves in closed and
  open networks via deep reinforcement learning,'' in \emph{Proc. ITSC}, Maui,
  HI, USA, Nov. 2018.

\bibitem{shi2020efficient}
T.~Shi\emph{,~et~al.}, ``Efficient connected and automated driving system with
  multi-agent graph reinforcement learning,'' \emph{arXiv preprint
  arXiv:2007.02794}, 2020.

\bibitem{tan1993multi}
M.~Tan, ``Multi-agent reinforcement learning: Independent vs. cooperative
  agents,'' in \emph{Proc. Int. Conf. Mach. Learn.}, University of
  Massachusetts, Amherst, Jun. 1993.

\bibitem{xu2021gradient}
X.~Xu\emph{,~et~al.}, ``The gradient convergence bound of federated multi-agent
  reinforcement learning with efficient communication,'' \emph{arXiv preprint
  arXiv:2103.13026}, 2021.

\bibitem{foerster2016learning}
J.~Foerster\emph{,~et~al.}, ``Learning to communicate with deep multi-agent
  reinforcement learning,'' in \emph{Proc. NeurIPS}, Barcelona, Spain, Dec.
  2016.

\bibitem{jiang2018learning}
J.~Jiang\emph{,~et~al.}, ``Learning attentional communication for multi-agent
  cooperation,'' in \emph{Proc. NeurIPS}, Montréal, Canada, Dec. 2018.

\bibitem{kim2021communication}
W.~Kim\emph{,~et~al.}, ``Communication in multi-agent reinforcement learning:
  {I}ntention sharing,'' in \emph{Int. Conf. Learn. Represent.}, Virtual,
  Online, May 2021.

\bibitem{xu2022trustable}
X.~Xu\emph{,~et~al.}, ``Trustable {P}olicy {C}ollaboration {S}cheme for
  {M}ulti-{A}gent {S}tigmergic {R}einforcement {L}earning,'' \emph{IEEE Commun.
  Lett.}, vol.~26, no.~4, pp. 823--827, Apr. 2022.

\bibitem{barbieri2022communication}
L.~Barbieri\emph{,~et~al.}, ``Communication-efficient {D}istributed {L}earning
  in {V}2{X} {N}etworks: {P}arameter {S}election and {Q}uantization,'' in
  \emph{Proc. IEEE Globecom}, Rio de Janeiro, Brazil, Dec. 2022.

\bibitem{hu2019decentralized}
C.~Hu\emph{,~et~al.}, ``Decentralized federated learning: A segmented gossip
  approach,'' \emph{arXiv preprint arXiv:1908.07782}, 2019.

\bibitem{kairouz2021advances}
P.~Kairouz\emph{,~et~al.}, ``Advances and open problems in federated
  learning,'' \emph{Found. Trends Mach. Learn.}, vol.~14, no. 1--2, pp. 1--210,
  Jun. 2021.

\bibitem{schulman2017proximal}
J.~Schulman\emph{,~et~al.}, ``Proximal policy optimization algorithms,''
  \emph{arXiv preprint arXiv:1707.06347}, 2017.

\bibitem{hegedHus2019gossip}
I.~Heged{\H{u}}s\emph{,~et~al.}, ``Gossip learning as a decentralized
  alternative to federated learning,'' in \emph{19th IFIP WG 6.1 International
  Conference on Distributed Applications and Interoperable Systems}, Kongens
  Lyngby, Denmark, Jun. 2019.

\bibitem{vinitsky2018benchmarks}
E.~Vinitsky\emph{,~et~al.}, ``Benchmarks for reinforcement learning in
  mixed-autonomy traffic,'' in \emph{Proc. CoRL}, Zürich, Switzerland, Oct.
  2018.

\bibitem{treiber2000congested}
M.~Treiber\emph{,~et~al.}, ``Congested traffic states in empirical observations
  and microscopic simulations,'' \emph{Physical review E}, vol.~62, no.~2, p.
  1805, Aug. 2000.

\bibitem{xiao_stochastic_2023}
B.~Xiao\emph{,~et~al.}, ``Stochastic graph neural network-based value
  decomposition for multi-agent reinforcement learning in urban traffic
  control,'' in \emph{Proc. {{IEEE VTC}} 2023-{{Spring}}}, {Florence, Italy},
  Jun. 2023.

\end{thebibliography}
\end{document}